\documentclass[a4paper]{jpconf}
\usepackage{graphicx}

\usepackage{bm}
\usepackage{amsfonts}
\usepackage{amsmath, amsthm, amssymb}
\usepackage{color}
\newcommand{\gtappr}{{{\lower4pt\hbox{$>$} } \atop \widetilde{ \ \ \ }}}
\newcommand{\ltappr}{{{\lower4pt\hbox{$<$} } \atop \widetilde{ \ \ \ }}}

\newcommand{\beq}{\begin{equation}}
\newcommand{\eeq}{\end{equation}}

\newcommand{\ie}{\textit{i.e. }}
                            
\newcommand{\ybal}{$\beta$-YbAlB$_4\,$}
\newcommand{\aybal}{$\alpha$-YbAlB$_4\,$}

\newsavebox{\fmbox}

\begin{document}
\title{Anisotropic transverse magnetoresistivity in \aybal}

\author{Yosuke Matsumoto, Jinpyo Hong, Kentaro Kuga, Satoru Nakatsuji}

\address{Institute for Solid State Physics, University of Tokyo, Kashiwa 277-8581, Japan}

\ead{matsumoto@issp.u-tokyo.ac.jp}

\begin{abstract}
We measured the transverse magnetoresistivity of the mixed valence compound \aybal. 
Two configurations were used where   
current was applied along [110] direction for both and magnetic field was applied along [-110] and $c$-axis. 
We found the transverse magnetoresistivity is highly anisotropic. In the weak field below 1 T, it is consistent with 
stronger $c$-$f$ hybridization in the $ab$ plane which was suggested from the previous zero field resistivity measurements. 
At the higher field above 3 T, we observed a negative transverse magnetoresistivity for the field applied along the $c$-axis. 
The temperature dependences of the resistivity measured at several different fields suggest the suppression of the heavy fermion behavior at the characteristic field 
of $\sim 5.5$ T.
\end{abstract}

\section{Introduction}
Quantum criticality (QC) in heavy fermion systems has been studied extensively for the past few decades. 
To date, the most of the studies have been restricted to the Kondo lattice systems with integer valence where 
a quantum critical point (QCP) is usually found on the border of magnetism. 
On the other hand, there is a growing attention to the possibility of a novel QC beyond the conventional understanding based on 
the spin-density-wave type instability\cite{Lohneysen07, gegenwart08}.  
Among them, the first Yb-based heavy fermion superconductor \ybal provides a unique example of a QC 
in the strongly mixed valence state \cite{nakatsuji08, KugaPRL, matsumoto-ZFQCP, ybal-valency}. 
Indeed, the QC cannot be described by the standard theory for the spin-density-wave instability\cite{Hertz76, Moriya85, Millis93}.  
The diverging magnetic susceptibility along the $c$-axis exhibits the $T/B$ scaling 
in the wide temperature ($T$) and magnetic field ($B$) region spanning 3 $\sim$ 4 orders of magnitude\cite{matsumoto-ZFQCP}. 
This indicates that the QC emerges without tuning any control parameter, suggesting a formation of an anomalous metallic phase. 

\ybal has the locally isostructral polymorph \aybal, which is also strongly mixed valent.  
The Yb valence estimated by a hard x-ray photoemission spectroscopy is +2.73 
for \aybal and +2.75 for \ybal at 20 K \cite{ybal-valency}.  
The valence fluctuation temperature scale was estimated to be $\sim 200$ - 300 K for both compounds\cite{matsumoto-ZFQCP, matsumoto-PRB84, matsuda-X-ray-JKPS62}.  
Surprisingly, these two systems exhibit a heavy fermion (HF) behavior with a characteristic temperature scale of $\sim 8$ K, 
which is far lower than the valence fluctuation scale \cite{matsumoto-ZFQCP}.  
This is quite unusual because Pauli paramagnetism is usually expected in the mixed valence compounds below the valence fluctuation temperature scale. 
The small temperature scale of $\sim 8$ K for the anomalous HF state may indicate that \aybal is also close to a QCP although 
it has a Fermi liquid (FL) ground state at zero field in contrast to \ybal \cite{matsumoto-PRB84}. 
Recently, it is suggested that the HF behavior is suppressed under the field above $\sim$5.5 T in both compounds\cite{matsumoto-SCES2013, matsumoto-betafull}. 

Another remarkable feature for both systems is the anisotropic hybridization between conduction and $f$ electrons ($c$-$f$ hybridization). 
From the argument based on the 
local symmetry of the Yb site, the crystal field ground doublet of both $\alpha$- and \ybal is suggested to be made solely of $|J_z = \pm 5/2>$~\cite{Andriy09}. 
In this case, the $c$-$f$ hybridization is expected to be highly anisotropic 
and have a node along the $c$-axis\cite{ramires-PRL109}. 
Indeed, it was already suggested experimentally from the resistivity measurements of \aybal \cite{matsumoto-PRB84}. 
The resistivity is highly anisotropic and the one in the $ab$-plane is 
10 times larger than the one along the $c$-axis, which is consistent with the hybridization node along the $c$-axis.  
Interestingly, it has also been pointed out that the anisotropic $c$-$f$ hybridization plays an important role in the formation of 
HF state under the strong valence fluctuation and the novel QC found in \ybal \cite{eoin-PRL109, matsumoto-betafull, ramires-PRL109}.  

In order to further examine the possibility of the anisotropic $c$-$f$ hybridization, here, we measured the transverse magnetoresistivity (TMR) of \aybal for the 
current ($I$) along [110] direction at low temperatures below 1 K. 
We found the TMR is highly anisotropic. The one in the field applied along the $c$-axis is quite different from the one in the field applied along the $ab$-plane. 
Furthermore, the anisotropy in the weak field below 1 T is consistent with the stronger $c$-$f$ hybridization in the $ab$-plane. 
On the other hand, we observed a negative TMR for the field applied along the $c$-axis above $\sim 3$ T, 
which corresponds to the suppression of the HF behavior mentioned above. 
The suppression of the HF behavior 
was also observed in the temperature dependence of the resistivity measured at several different fields.

We used a high purity single crystal of \aybal with RRR (Residual Resistivity Ratio) $\sim$ 20 grown by a flux method \cite{Macaluso07}. 
It was reshaped to a size of $\sim$ 0.5 mm in [110] direction and $\sim$ 20 $\mu$m $\times \sim$ 20 $\mu$m in 
its perpendicular direction. 
The resistivity and TMR measurements were made by the conventional AC four-terminal method.

\section{Results and discussion} 
\begin{figure}[bt]
\begin{center}
\includegraphics[width=36pc]{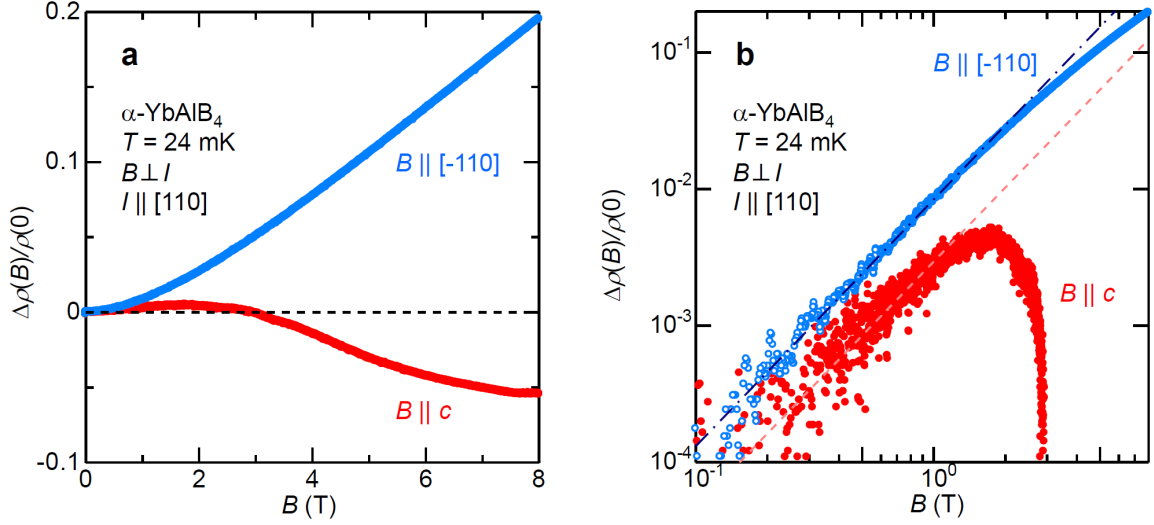}
\end{center}
\caption{\label{TMR} (a) Normalized transverse magnetoresistivity $\Delta\rho (B)/\rho (0)$ of \aybal measured at $T=$ 24 mK for $B || [-110]$ and 
$B || c$. Current is applied along [110] for both. Here, $\rho(0)$ is a zero-field resistivity at $T=$ 24 mK and  $\Delta\rho (B) = \rho (B)-\rho (0)$. 
(b) $\Delta\rho (B)/\rho (0)$ in full logarithmic scale.  
The long dashed short dashed line represents a power law fit for $B || [-110]$ below 1.5 T, which indicates $\Delta\rho (B)/\rho (0)\propto B^{1.8}$ 
behavior. 
The dashed line represents a power law fit with a power fixed to 1.8 for $B || c$ below 0.9 T. See text for detail. 
}
\end{figure}

First we present in Fig. \ref{TMR} (a) the field dependence of the TMR  of \aybal measured at 24 mK for the two configurations described above.  
Here, TMR, which is expressed as $\Delta\rho (B)$, is normalized by the zero-field resistivity $\rho (0)$.
As clearly seen from the figure, TMR is highly anisotropic. While TMR for the field along the [-110] direction shows a
monotonic increase, the one for the $c$-axis exhibits negative magnetoresistivity at field above $\sim 3$ T. 
In order to see the low field region in detail, we plot the full logarithmic version of TMR versus $B$ in Fig. \ref{TMR} (b), 
where we find a clear power law behavior of $\Delta\rho (B)/\rho (0)\propto B^{1.8}$ for $B ||$ [-110] up to $\sim 1.5$ T. 
On the other hand, TMR for $B || c$ is smaller and it is rather hard to see if there is a power law behavior in the low field limit.  

In the low field, TMR is expected to be proportional to $(m^*)^{-2}$ where $m^*$ is 
a (cyclotron) effective mass\cite{Kittel63}. 
TMR arises from a cyclotron motion of electrons which is perpendicular to the applied magnetic field. 
Therefore, while TMR for $B || c$ reflects the motion within the $ab$-plane, TMR for $B || ab$ reflects the motion 
both within the $ab$-plane and along the $c$-axis. 
If we assume that $m^*$ for the motion within the $ab$-plane ( $m^*_{ab}$) and $m^*$ along the $c$-axis ($m^*_{c}$) are different, 
then, TMR will be also anisotropic depending on the field direction. 
Assuming $(\Delta\rho (B))_{B || c}\propto (m^*_{ab})^{-2}$ and  $(\Delta\rho (B))_{B || ab}\propto (m^*_{ab}m^*_{c})^{-1}$, 
the ratio between the two will be given by 
\beq 
\frac{(\Delta\rho (B))_{B || ab}}{(\Delta\rho (B))_{B || c}} \sim \frac{m^*_{ab}}{m^*_{c}}. 
\label{eq_TMR} 
\eeq 
In our results, TMR is larger for $B || ab$ in the weak field. This indicates that $m^*_{ab}$ is larger than $m^*_{c}$ according to the above equation, 
consistent with the stronger $c$-$f$ hybridization within the $ab$-plane.  

To estimate the ratio between $m^*_{ab}$ and $m^*_{c}$, 
we also made a power law fit to TMR for $B || c$ below 0.9 T as shown in Fig. \ref{TMR} (b). Here we assumed $\Delta\rho (B)/\rho (0)\propto B^{1.8}$.    
By comparing the coefficients of the $B^{1.8}$ behavior for both $B || ab$ and $B || c$,  
the ratio in the low field below 0.9 T was roughly estimated to be $m^*_{ab}/m^*_{c} \sim 2.9$. 
On the other hand, another estimation is available from $A$ coefficient of the $T^2$ temperature dependence of the resistivity 
defined as $\rho = \rho _0 + AT^2$. Here, $\rho _0$ is the resistivity at zero temperature limit (residual resistivity). 
As already discussed in the previous work\cite{matsumoto-PRB84}, $A$ coefficient is highly anisotropic, \ie, the 
one for the current applied along the $ab$-plane ($A_{ab}$) is 13 times larger that the one along the $c$-axis ($A_c$).  
Using $A\propto (m^*)^2$, the anisotropy in the effective mass is roughly estimated to be 
$m^*_{ab}/m^*_{c} = \sqrt{A_{ab}/A_{c}} \sim 3.7$, which is of the same order as the above estimate from TMR. 
Note that, 
as we will discuss later, the temperature dependence of the resistivity along the $ab$-plane is slightly 
different from $T^2$ behavior expected for the FL. Instead, it exhibits $T^{1.8}$ dependence as 
it was already pointed out in the previous work\cite{matsumoto-PRB84}. 
Strictly speaking, this indicates that the $A$ coefficient can not be defined for the resistivity along the $ab$-plane.   
Nevertheless, in order to estimate  $m^*_{ab}/m^*_{c}$, 
here we tentatively fixed the exponent to 2.0, the same value as in the determination of $A_c$. 

The negative TMR found for $B || c$ above $\sim 3$ T 
may correspond to the suppression of the HF behavior. 
We observed the suppression behavior also in the temperature dependences of the resistivity measured at several different fields 
in the same current and field configurations as TMR measurements. 
The measurements were done at the temperature range $0.02 \leq  T \leq  0.7$ K. 

\begin{figure}[bt]
\begin{center}
\includegraphics[width=36pc]{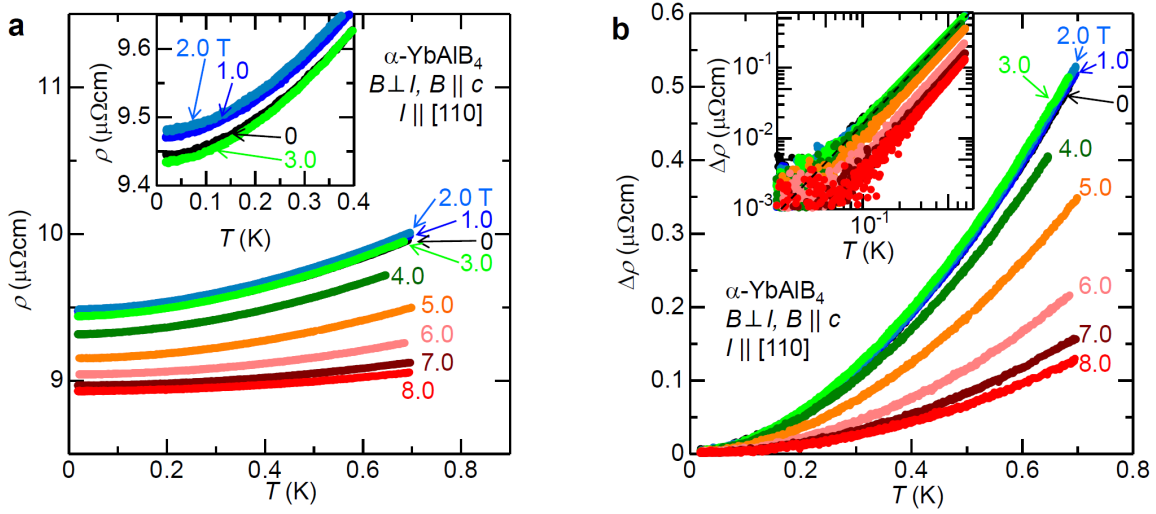}
\end{center}
\caption{\label{rhoab_vs_Bc}(a) Temperature dependence of the resistivity with the current along [110] at various fields along the $c$-axis. 
Inset shows the enlargement of the low temperature part of the data at $B = 3.0$ T or lower fields. (b) Temperature dependence of 
$\Delta\rho$ for the same data as in (a). Here $\Delta\rho = \rho - \rho _0$ and $\rho _0$ is the residual resistivity. 
Inset shows the full logarithmic plot of the main figure. The dashed line indicates  $\Delta\rho\propto T^{1.8}$ behavior. 
}
\end{figure}

Figure \ref{rhoab_vs_Bc} (a) shows the temperature dependence of the resistivity with $I ||$ [110] and $B || c$. 
Corresponding to TMR discussed above, by applying magnetic field, it increases slightly up to 2 T and 
decreases above 3 T. If we subtract $\rho_0$ from the data, Figure \ref{rhoab_vs_Bc} (b) is obtained, where we note that 
the temperature dependence is significantly suppressed above 3 T while it does not indicate almost any change up to this field. 
As we will discuss later, this corresponds to the suppression of the HF behavior. 
The temperature dependence indicates the power law behavior $\Delta\rho\propto T^{1.8}$ at low field as shown in the inset of 
Fig. \ref{rhoab_vs_Bc} (b). 

\begin{figure}[bt]
\begin{center}
\includegraphics[width=36pc, clip]{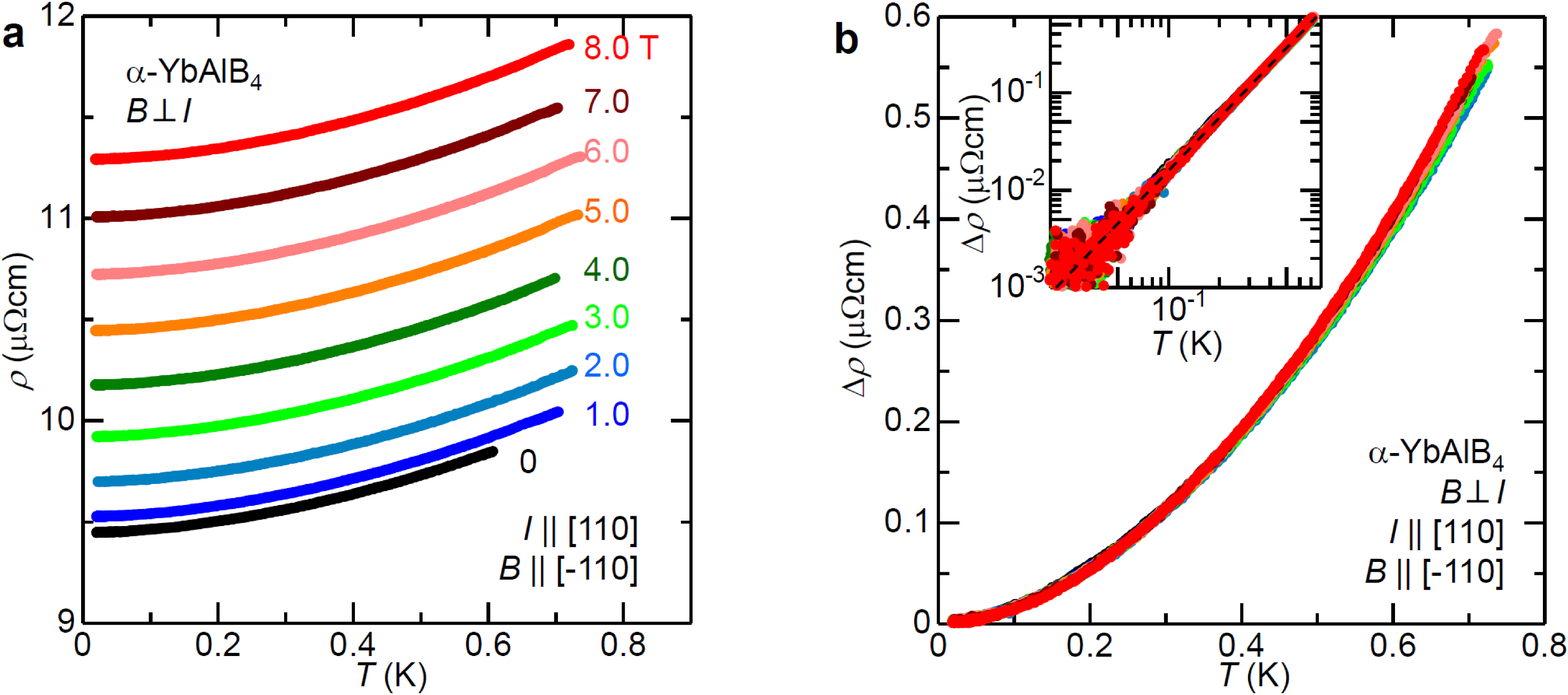}
\end{center}
\caption{\label{rhoab_vs_Bab}(a) Temperature dependence of the resistivity with the current along [110] at various fields along [-110] direction. 
(b) Temperature dependence of 
$\Delta\rho$ for the same data as (a). Here $\Delta\rho = \rho - \rho _0$ and $\rho _0$ is the residual resistivity. 
Inset shows the full logarithmic plot of the main figure.  
}
\end{figure}

The temperature dependences of the resistivity measured with  $I ||$ [110] and $B ||$ [-110] are shown in 
Fig. \ref{rhoab_vs_Bab} (a) and (b). By applying magnetic field, the isothermal resistivity increases due to TMR. 
If we compare $\Delta\rho$ obtained after subtracting $\rho _0$, they overlap quite well to each other up to $B=8$ T.  
Therefore, there is no suppression of the HF behavior for $B ||$ [-110], which is consistent with the Ising anisotropy of the system. 
They exhibit the power law behavior $\Delta\rho\propto T^{1.8}$ up to $B=$ 8 T as shown in the inset of 
Fig. \ref{rhoab_vs_Bab} (b).  

\begin{figure}[bt]
\includegraphics[width=17pc]{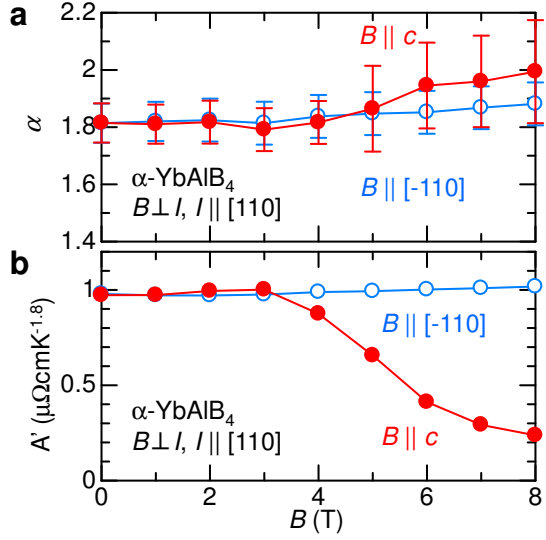}\hspace{2pc}%
\begin{minipage}[b]{18pc}\caption{\label{exponent_Acoeff} 
 Magnetic field dependence of 
(a) the resistivity exponent $\alpha$ defined by $\Delta\rho = AT^{\alpha}$ and 
(b) the coefficient $A^{'}$ for $T^{1.8}$ behavior of $\Delta\rho$, which is defined as $\Delta \rho = A^{'} T^{1.8}$ 
for current applied along the [110] direction. 
Field is applied along the $c$-axis (closed circles) and the [-110] direction (open circles). See text for detail. 

}
\end{minipage}
\end{figure}

In order to discuss the temperature dependence in detail, we estimated the exponent $\alpha$ defined by 
$\Delta\rho = A_{\alpha}T^{\alpha}$ at each field. The definition of $\alpha$ gives $\alpha = \partial \log \Delta\rho / \partial \log T$ as a temperature dependent quantity.  
We found that $\alpha$ is almost temperature independent up to $\sim 0.7$ K for all the field up to 8 T in both $B || c$ and $B ||$ [-110]. 
The field dependences of $\alpha$ obtained at 0.1 K for each field direction are shown in Fig. \ref{exponent_Acoeff}(a). 
The error bars are mostly coming from the errors in $\rho _0$ for each data.   
For $B ||$ [-110], $\alpha$ is almost field independent with a value $\sim$ 1.8. 
On the other hand, it increases a little bit towards the normal value of 2.0 above 5 T for $B || c$. 
This may indicate that the ground state of \aybal is slightly deviating from the FL at zero-field and 
recovers the FL state after the suppression of the HF behavior above 5 T along the $c$-axis.

The suppression of the HF behavior is clearly seen in Fig. \ref{exponent_Acoeff} (b). 
Here, we estimated a coefficient $A'$ defined by $\Delta\rho = A'T^{1.8}$ because 
$\Delta\rho\propto T^{1.8}$ behavior is always observed except for the slight increase of the exponent $\alpha$ above 5 T for $B || c$. 
Although the definition is different from the one for FL, $A'$ is still expected to give an estimate of $m^*$.
Note that the same field evolution is obtained even if we plot $A$ defined by $\Delta\rho = AT^{2}$. 
The field evolution of $A'$ indicates the suppression of the HF behavior above a field scale of $\sim$5.5 T, characterized by the inflection point.

\section{Conclusion}
We measured the transverse magnetoresistivity (TMR) in \aybal for the 
current ($I$) along [110] direction at low temperatures below 1 K. 
We found the TMR is highly anisotropic and those in the field applied along $c$-axis and in the $ab$-plane are quite different to each other. 
The anisotropy in the weak field below 1 T is consistent with the stronger $c$-$f$ hybridization in the $ab$-plane. 
Furthermore, we observed a negative TMR for the field applied along the $c$-axis above $\sim 3$ T, 
which should arise from the suppression of the HF behavior by magnetic field. 
The temperature dependence of the resistivity also indicates the suppression of the HF behavior with a characteristic field scale of $\sim$ 5.5 T. 

\ack
We thank E. C. T. O'Farrell, P. Coleman, A. H. Nevidomskyy, 
S. Watanabe, C. Broholm, K. Ueda 
for supports and 
useful discussions.
This work is partially supported by Grants-in-Aid (No.
21684019, 25707030) from JSPS, by
Grants-in-Aids for Scientific Research on Innovative Areas (No. 20102007, No. 21102507) from MEXT, Japan.

\section*{References}
\providecommand{\newblock}{}

\end{document}